\documentclass[3p]{elsarticle}
\usepackage{amsmath,amssymb}
\usepackage{lineno}
\usepackage[bookmarks=false]{hyperref}
\usepackage{threeparttable}

\begin{document}

\begin{frontmatter}

\title{Atomic masses of intermediate-mass neutron-deficient nuclei with \\  relative uncertainty down to 35-ppb via multireflection time-of-flight mass spectrograph}

\author[tsukuba,riken]{S. Kimura}\corref{correspondingauthor}
\cortext[correspondingauthor]{Corresponding author}
\ead{sota.kimura@riken.jp}

\author[riken]{Y.~Ito\fnref{footnote}}
\author[riken]{D.~Kaji}
\author[kek]{P.~Schury}
\author[kek,riken]{M.~Wada}
\author[riken]{H.~Haba}
\author[ibs]{T.~Hashimoto}
\author[kek]{Y.~Hirayama}
\author[in2p3]{M.~MacCormick}
\author[kek]{H.~Miyatake}
\author[ibs]{J.~Y.~Moon}
\author[riken]{K.~Morimoto}
\author[tsukuba,riken,kek]{M.~Mukai}
\author[in2p3]{I.~Murray}
\author[tsukuba]{A.~Ozawa}
\author[riken]{M.~Rosenbusch}
\author[msu]{H.~Schatz}
\author[riken]{A.~Takamine}
\author[riken,kyusyu]{T.~Tanaka}
\author[kek]{Y.~X.~Watanabe}
\author[nmsu]{H.~Wollnik}

\address[tsukuba]{Department of Physics, University of Tsukuba, Ibaraki 305-8577, Japan}
\address[riken]{Nishina Center for Accelerator Based Science, RIKEN, Saitama, 351-0198, Japan}
\address[kek]{Wako Nuclear Science Center (WNSC), Institute of Particle and Nuclear Studies (IPNS), \\ High Energy Accelerator Research Organization(KEK), Saitama, 351-0198, Japan}
\address[ibs]{Rare Isotope Science Project (RISP), Institute of Basic Science (IBS), Daejeon 305-811 Korea}
\address[in2p3]{Institut de Physique Nucl$\acute{e}$aire, IN2P3-CNRS, Universit$\acute{e}$ Paris-Sud, Universit$\acute{e}$ Paris-Saclay, 91406 Orsay Cedex, France}
\address[msu]{Department of Physics and Astronomy and National Superconducting Cyclotron Laboratory, Michigan State University, East Lansing, Michigan 48824 USA}
\address[kyusyu]{Department of Physics, Kyushu University, Hakozaki, Higashi-ku, Fukuoka 812-8581, Japan}
\address[nmsu]{New Mexico State University, Las Cruces, New Mexico 88001, USA}

\fntext[footnote]{Present address: Department of Physics, McGill University, Quebec, H3A 2T8, Canada}

\begin{abstract}
High-precision  mass measurements of $^{63}$Cu, $^{64-66}$Zn, $^{65}$Ga, $^{65-67}$Ge, $^{67}$As, $^{78,81}$Br, $^{80}$Rb, and $^{79}$Sr  were performed utilizing a multireflection time-of-flight mass spectrograph combined with the gas-filled recoil ion separator GARIS-II. In the case of $^{65}$Ga, a mass uncertainty of 2.1~keV, corresponding to a relative precision of $\delta m / m = 3.5\times10^{-8}$, was obtained and the mass value is in excellent agreement with the 2016 Atomic Mass Evaluation.  For $^{67}$Ge and $^{81}$Br, where masses were previously deduced through indirect measurements, discrepancies with literature values were found.  The feasibility of using this device for mass measurements of nuclides more neutron-deficient side, which have significant impact on the $rp$-process pathway, is discussed.
\end{abstract}


\end{frontmatter}


\section{Introduction}

Nuclear masses of nuclei along the $N=Z$ line are crucial in determining the rapid proton-capture ($rp$-) process pathway which drives explosive astronomical phenomena called type I X-ray bursts \citep{Parikh2013}. The $rp$-process follows a pathway which transits through several key nuclei, notably $^{64}$Ge, $^{68}$Se and, $^{72}$Kr, the exact route being strongly dependent on the effective lifetimes of these waiting-point nuclei. Effective lifetimes of waiting-point nuclei depend exponentially on the $Q$-values of the one-(two-)proton capture reaction $Q_{p}$ ( $Q_{2p}$) at an environmental temperature lower (higher) than $\sim 1.4$~GK \citep{Brown2002}. The uncertainties in the nuclear masses needed to determine these $Q$-values should be less than $\sim$10~keV to significantly reduce the uncertainties of $rp$-process calculations \citep{Schatz2006,Schatz2013}. 

The masses of the waiting-point nuclei, $^{64}$Ge \citep{Schury2007}, $^{68}$Se \citep{Savory2009}, and $^{72}$Kr \citep{Rodriguez2004} have been measured with precisions beyond that required for X-ray burst studies.  However, among the six nuclides ($^{65}$As, $^{69}$Br, $^{73}$Rb, $^{66}$Se, $^{70}$Kr, and $^{74}$Sr ) which are the counterparts for calculating $Q_{p}$- and $Q_{2p}$-values, only the masses of $^{65}$As and $^{69}$Br have been determined experimentally, with uncertainties of 85~keV \citep{Tu2011} and 40~keV \citep{Rogers2011}, respectively. For the others, only theoretical predictions are given. A recent $Q$-value sensitivity study pointed out that the $^{65}$As mass uncertainty has significant impact on the light curves and the ash compositions of X-ray bursts \citep{Schatz2017}. 

High-precision experimental mass data of nuclides near the $N=Z$ line are also necessary for verification of the Standard Model through the unitarity of the Cabibbo-Kobayashi-Masukawa (CKM) matrix. The ``corrected" ${\cal F}t$-values of super allowed $0^{+} \rightarrow 0^{+}$ $\beta^{+}$-decay between $T = 1$ analog states are directly related to the dominant term in the top-row sum of the CKM matrix \citep{Towner2010}. To calculate the ${\cal F}t$-values, the necessary nuclear parameters are partial lifetimes of the $0^{+} \rightarrow 0^{+}$ transition  and the corresponding $Q_{\rm EC}$-values. For nuclear masses, a relative precision of $\delta m/m \lesssim 5 \times 10 ^{-8}$ is required. The unitarity of the CKM matrix is confirmed to the level of $1.2 \times 10^{-4}$ with the uncertainties of present nuclear data \citep{Hardy2015}. A large fraction of the nuclear uncertainty stems from the mass uncertainty of $^{66}$As (30~keV \citep{Schury2007}) and the ambiguity in the $^{70}$Br mass value \citep{Hardy2015, Savory2009}.

Half-lives of these as-yet insufficiently studied nuclides, excluding highly proton-unbound $^{69}$Br and $^{73}$Rb, span from tens to hundreds of milliseconds. These short half-lives make them difficult to effectively measure with Penning traps.  In contrast, the multireflection time-of-flight mass spectrograph (MRTOF-MS) has advantages, discussed below. 

The MRTOF-MS has been developed for both mass measurement and isobar separation techniques in recent years. Typical mass resolving power of MRTOF-MS reaches $R_m > 10^5$ with short measurement times of less than 10 ms. MRTOF-MS have been applied for mass measurement of heavy nuclei \cite{Schury2017,Ito2017submitted,Rosenbusch2018submitted} and light neutron-rich nuclei \cite{Wienholtz2013,Rosenbusch2015,Leistenschneider2017} in several facilities. The achieved relative mass uncertainties have so far reached the order of $\delta m/m \sim 10^{-7}$, providing measurements relevant for nuclear-structure physics and some astrophysics. However, the measurement precision reported so far from on-line MRTOF-MS measurements is not sufficient for $rp$-process and unitarity of CKM matrix. 

The SHE-mass facility at RIKEN has been constructed to enable mass measurements of super-heavy elements (SHE) \citep{Schury2017}. However, the applicable mass region is not limited and can be extended to light nuclei as required for studies relevant to the $rp$-process. The facility consists of an MRTOF-MS \citep{Schury2014} coupled with the gas-filled recoil ion separator GARIS-II \citep{Kaji2013} via a cryogenic gas-cell and an ion transport system, and has been utilized for the measurement of neutron-deficient nuclei produced by symmetric fusion-evaporation reactions for the first time. Several medium-mass reaction products have been identified and measured with high precision. In the present study, we report the on-line measurements and discuss the data-analysis and achieved precision in the context of utilizing the SHE-mass facility for the most exotic cases in near future.

\section{Experiment}

\begin{figure*}[!t]
  \centering
  \includegraphics[width=0.9\textwidth, bb=0 0 842 595, clip, trim=0 45 0 40]{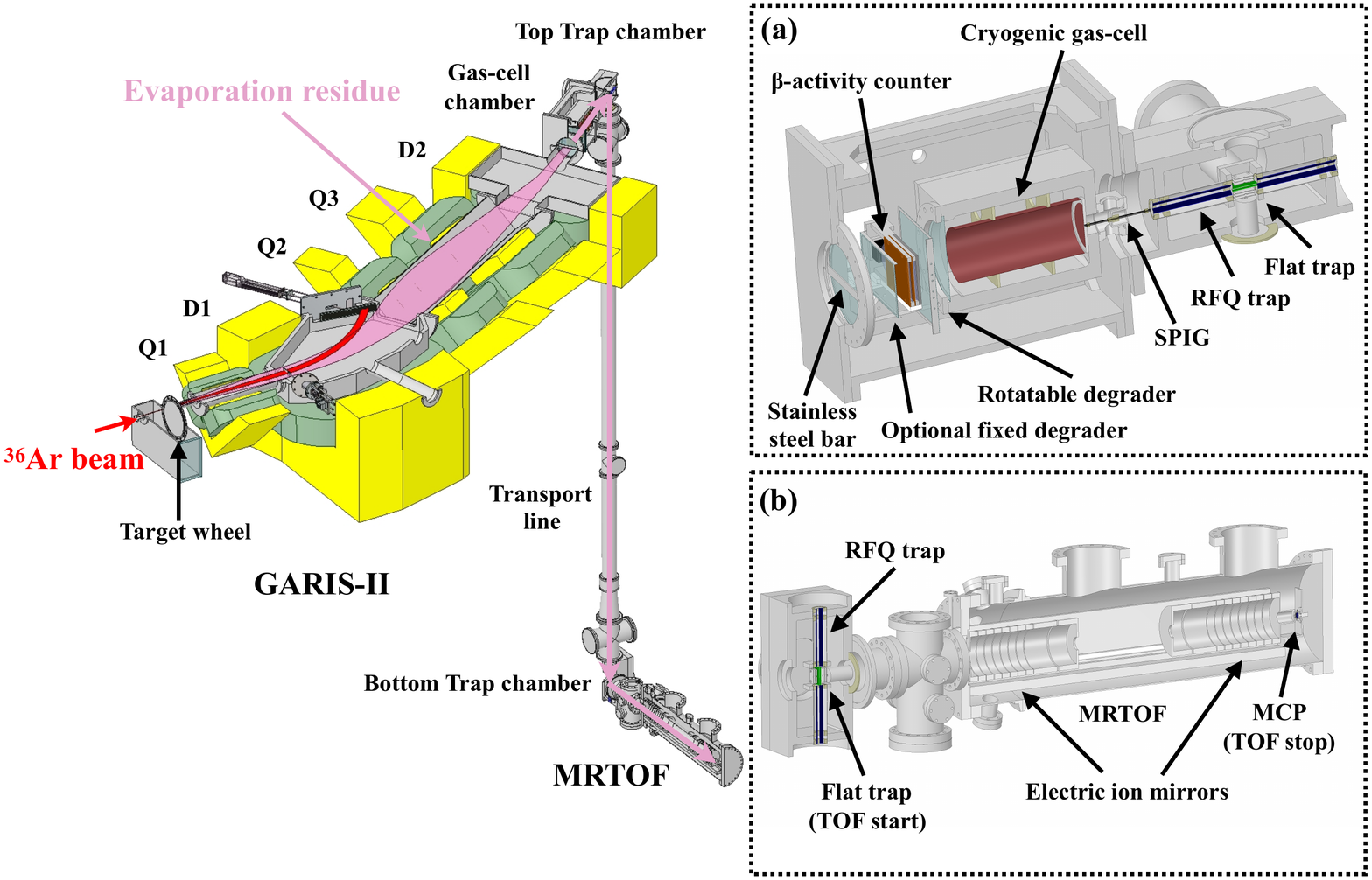}
  \caption{Overview of the SHE-mass facility. (a) Details of gas-cell and top trap chambers. (b) Details of bottom trap chamber and MRTOF device. See texts for details. \label{setup}}
\end{figure*}

A $3.30 \ {\rm MeV/nucleon}  \ ^{36}{\rm Ar}^{10+}$ beam, with maximum intensity of $ 3 \ {\rm p \mu A}$, was provided by the RIKEN linear accelerator RILAC. Sixteen ${\rm Mo}^{\rm nat}{\rm S}_2$ targets mounted on a 300~mm diameter rotating wheel \citep{Kaji2015} were employed to produce neutron-deficient  nuclei through fusion-evaporation reactions via $^{\rm nat}{\rm S}$($^{36}{\rm Ar}$,X). The ${\rm Mo}^{\rm nat}{\rm S}_2$ targets were prepared by a spray coating technique \citep{Greene2002} on 3~$\mu$m Ti backing foils. The average target thickness was $1.9 \ {\rm mg/cm^2}$. The wheel rotated at 2000~rpm during irradiation. The helium gas pressure of GARIS-II was set to 1~mbar during the measurements.

In the SHE-mass facility, a detector system tracking activities is specialized for (super-)heavy, $\alpha$-decaying nuclei. The modified setup in the GARIS-II focal plane for intermediate-mass $\beta$-decaying nuclei is shown in Fig.~\ref{setup}--(a). Since there is a lack of data on the mean charge states of high-velocity intermediate-mass nuclei in dilute helium gas, a retractable $\beta$-activity counter was installed at the GARIS-II focal plane for finding the optimum $B\rho$-value. It consisted of double-layered, 1~mm and 2~mm thick plastic scintillators with two 1.5~mm copper energy degraders to suppress the low-energy $\beta$-rays ($E_{\beta} \lesssim 4$~MeV).

A schematic view of the SHE-mass facility is shown in Fig.~\ref{setup}. Nuclei produced via fusion-evaporation reactions (pink band of Fig.~\ref{setup}) will be separated from unwanted light primary beam-like particles (red band of Fig.~\ref{setup}) and  guided to the focal plane of GARIS-II. When symmetric reactions are used, a difficulty arises from the lack of separation between the primary beams and the reaction products in GARIS-II due to their small difference in magnetic rigidity, $B\rho$. Thus, intense primary beams could degrade gas cell performance or even lead to breakage of the thin Mylar windows between GARIS-II and the gas cell. To overcome this, a system of adjustable beam stoppers has been developed to enhance the suppression of such particles (a detailed discussion will be provided in a forthcoming publication).  Additionally, a stainless steel bar was set in front of the GARIS-II bulkhead to further protect its Mylar window from beam bombardments.

The SHE-mass facility, located after GARIS-II, comprises a cryogenic gas cell, an ion transport/preparation section consisting of two ion-trap systems separated by a $\sim 3\,\mathrm{m}$ transport line, and the MRTOF-MS. The gas cell was cooled to 90~K and pressurized with helium gas at 100~mbar room temperature equivalent. Two independent Mylar energy degraders, a 2.5~$\mu$m optional fixed degrader and a 5~$\mu$m rotatable degrader providing an effective thickness ranging from 5~$\mu$m to 9~$\mu$m, allowed optimization of the ion stopping ranges in the gas cell.. After stopping in the gas cell, the reaction products were extracted with an rf-carpet using the traveling-wave technique \cite{Arai2014}.

After extraction from the gas cell, the ions were captured, cooled, and bunched by the radio-frequency (RF) trap system in the top trap chamber.  This system consists of two linear radio-frequency quadrupole (RFQ) Paul traps, one on either side of a central RF trap with a planar electrode geometry (flat trap) \cite{Schury2009}. A Bradbury-Neilson gate \cite{Brunner2012} installed in the transport line between the two traps allows for rejection of unwanted molecular ions that may be extracted from the gas cell with a mass resolving power up to $R_m \sim 100$. After recapturing and cooling the ions again, the trapping and cooling process in the bottom flat trap is the final preparation step for the mass measurement.

The time-of-flight measurement begins the moment ions are perpendicularly ejected from the bottom flat trap to travel toward the MRTOF-MS; the ejection trigger signal also serves as the time-to-digital conversion (TDC) start trigger.  Due to a combination of differences in cable lengths and inherent properties of the MOSFETs utilized in the trap ejection circuitry, there is a delay ($t_0$) between actual ion ejection and the TDC start trigger.  This delay has been directly measured with an oscilloscope as well as by using false TDC stop triggers associated with switching noise and determined to be $t_0=40 $~ns. Its uncertainty was evaluated to be less than 10~ns \citep{Schury2017}. 

The ions are captured in the MRTOF-MS by temporarily lowering the potential of the entrance mirror and are then reflect back and forth between the electrostatic ion mirrors.  After the ions have undergone a a specific number of reflections ( $\sim$225 laps, corresponding to a flight time of $t_{\rm meas.}\sim 6 \ {\rm ms}$ for the masses under study herein) the potential of the exit mirror is temporarily lowered to allow ions to leave and travel to the ion detector, where they make a time focus and thereby a mass resolving power of  $R_\mathrm{m}\sim$120,000 is achieved.  After stopping in the gas cell, the average radioactive ion will spend less than 30~ms traversing the system before reaching the detector after the MRTOF-MS.

\section{Analysis method}
\label{analysis}

The measured time-of-flight (TOF) value for an ion, having mass $m$ and charge $q$,  which undergoes $\ell$ laps in the MRTOF-MS device can be represented by 
\begin{equation}
t_{\rm obs} = (a + b \cdot \ell) \sqrt{m/q} + t_0
\label{eqObsTOF}
\end{equation}
where $t_0$ is a constant offset dependent on the measurement system, and $a$ and $b$ are constants related to the non-reflection flight path and the flight path between consecutive reflections, respectively. In order to determine the mass of analyte ions, two different methods -- single referencing \cite{Ito2013} and double referencing \cite{Wienholtz2013,Rosenbusch2015} -- have been employed. In the case of the double reference method, two references of well-known mass which have experienced the same flight length are required. In contrast, the single reference method needs only one reference mass, however the value of $t_0$ must be determined independently.  Uncertainty in $t_0$ can dominate the systematic mass uncertainty in the case of large fractional mass differences between reference and analyte, but becomes negligibly small in this method, when using an isobaric mass reference.  

The TOF values varied as a function of time due to thermal expansion of the MRTOF-MS device and minor instabilities in the high-voltage power supply system for mirror electrodes \citep{Schury2014}. These TOF drifts could be compensated by use of an isobaric reference species for each measurement. The TOF corrections were performed within each subset, obtained by dividing the raw data set into ${\cal N}$ parts. For ions in each subset $i$ the corrected TOF $t_{\rm corr,i}$ are calculated with the following relation:
\begin{equation}
t_{\rm corr,i} = t_{\rm raw,i} \left( \frac{t_{\rm raw}}{t_{\rm i}}\right),
\end{equation}
where $t_{\rm raw,i}$ is the uncorrected TOF of each ion in subset $i$, while $ t_{\rm raw}$ and $t_{\rm i}$ are the fitted TOF center of the full raw spectra and of the $i^{th}$ spectral subset, respectively, of the isobaric reference species. This correction method was applied to all the data presented here. For a typical case, the width of the $^{81}$Sr peak with 221 laps (see Fig.~\ref{A79-81} -- (c)), was reduced from $\Delta t_{\rm{FWHM}}$=29.67(9)~ns to $\Delta t_{\rm{FWHM}}$=28.63(9)~ns ($\sim$4\%) by applying this TOF correction technique. In the present experiment, improvements of peak width by TOF drift compensation is small due to the relatively short measurement duration of $\approx$1~hr, however for the high-precision measurements we desire it is still important, while future measurements of more neutron-deficient nuclei will require longer measurement durations and necessitate the compensation technique moreso. 

To determine the masses of observed nuclides, the single reference method was adopted. In this method $m_{\rm X}$, the ionic mass of nuclide X, is given by Eq.~\ref{eqSingleRef}:
\begin{equation}
m_{\rm X} = \rho^2 m_{\rm ref} = \left( \frac{t_{\rm X}-t_0}{t_{\rm ref}-t_0} \right)^2 m_{\rm ref},
\label{eqSingleRef}
\end{equation}
where $\rho$ is the TOF ratio, $t_{\rm X}$ and $t_{\rm ref}$ are the TOF of nuclide X and the reference ion, respectively, $m_{\rm ref}$ is the mass of the reference ion,  $t_0$ is the constant time offset within the measurement system mentioned above,  and $q_{\rm X} = q_{\rm ref}$ is assumed. The systematic mass error resulting from the uncertainty of the constant time offset $\delta t_0$ can be estimated as follows \citep{Itothesis},
\begin{eqnarray}
\left( \frac{\delta m}{m} \right)_{\rm sys}  &=& \frac{2m_{\rm ref}}{m_{\rm X}} \frac{t_{\rm X}(t_{\rm X}-t_{\rm ref})}{t_{\rm ref}^3}\delta t_0
 \nonumber \\
&\approx& \frac{2\delta t_0}{t_{\rm ref}} \left(1- \sqrt{\frac{m_{\rm ref}}{m_{\rm X}}} \right).
\end{eqnarray}
When employing isobaric reference masses, one can calculate the relative systematic mass error induced by $\delta t_0$ to be on the order of $10^{-10}$, a negligible contribution in the present study.

In the present measurements, only singly-charged ions were measured. Thus, the atomic mass of nuclide X, $M_{\rm X}$, is given by 
\begin{equation}
M_{\rm X} = \rho^2 (M_{\rm ref} - m_{\rm e}) + m_{\rm e},
\end{equation}
where $M_{\rm ref}$ and $m_{\rm e}$ are the atomic mass of the reference nuclide and the electron rest mass, respectively. For mass values of the references, the 2016 Atomic Mass Evaluation (AME16) values \citep{Huang2016,Wang2016} were adopted. 

Excluding low statistics species which have less than a few hundreds events, the least squares fitting routine of the ROOT package \citep{Brun1997} was used to determine $\rho$. For the low statistics cases the maximum likelihood method was used.

Measurements were performed with two or more different number of laps for each A/q series. The final results of ${\cal N}$ measurements belonging to the same A/q series were obtained as their weighted average,
\begin{equation}
\overline{\rho^2} = \frac{\sum_{i=1}^{\cal N} w^2_i \rho^2_i}{\sum_{i=1}^{\cal N} w^2_i},
\end{equation}
where $w_i$ is a weight of each measurement and is defined by $w_i \equiv 1/ \delta (\rho_i^2)$.

\section{Fitting function}

\begin{figure}[h]
  \centering
   \includegraphics[width=0.5\textwidth, bb=0 0 842 595, clip, trim=35 30 35 80]{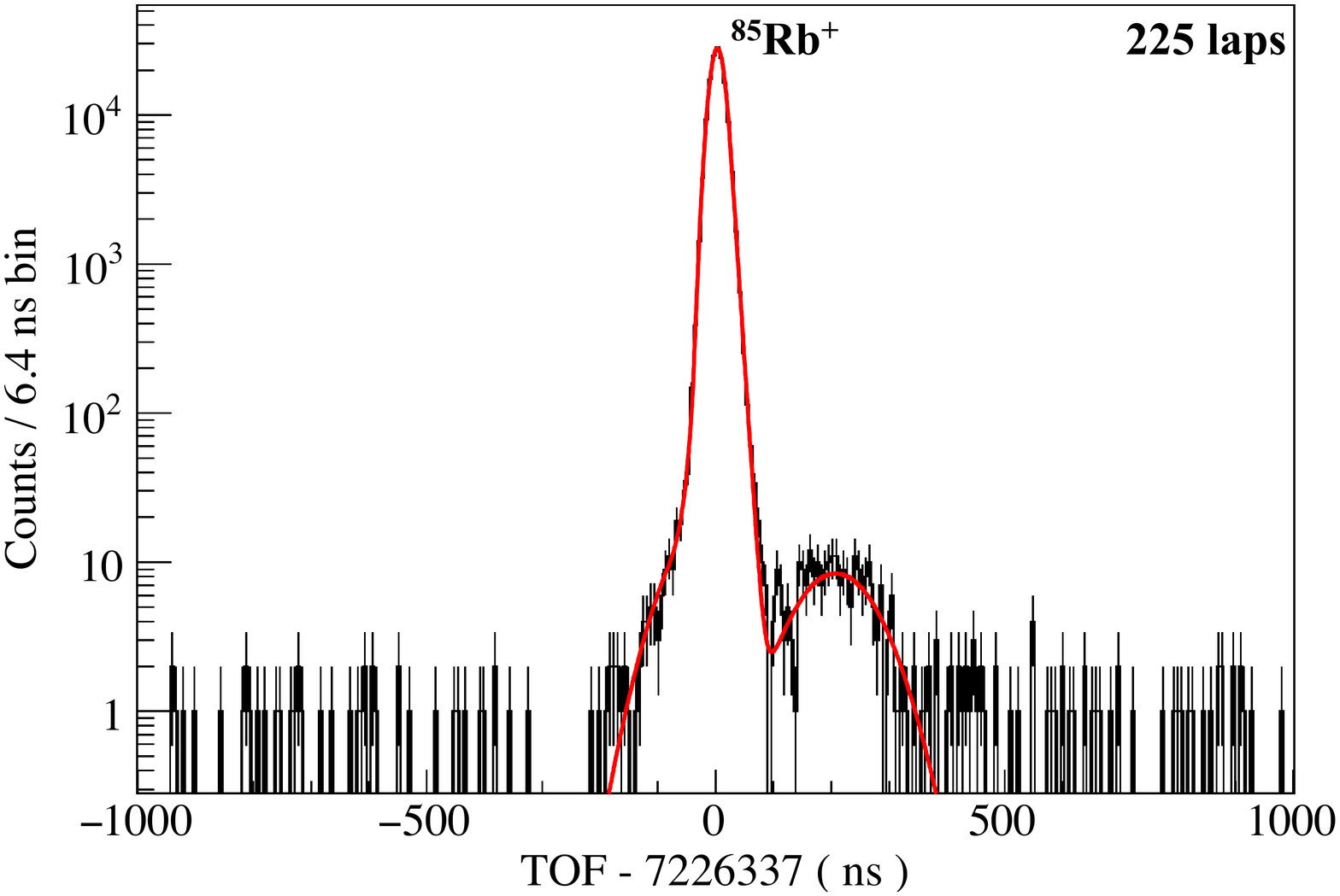}
  \caption{(Color online). Typical time-of-flight spectrum of $^{85}$Rb$^{+}$ ions. Such spectra were used to determine the fitting function, indicated by a red line, based on an exponential-Gaussian hybrid function. \label{RF}}
  \vspace*{-15pt}
\end{figure}

A phenomenological fitting function, based on an exponential-Gaussian hybrid function \citep{Ito2013,Schury2014,Lan2001}, was developed through the study of high-statistical offline measurements of $^{85}$Rb$^{+}$ ions. A typical TOF spectrum of $^{85}$Rb$^{+}$ ions is shown in Fig.~\ref{RF}. In order to reproduce the shape of main peak, we employed the function:
\begin{align}
f_{\rm p}(\tau) = \left\{
\begin{array}{l}
\mu \exp \left[ - \tau^2 / \nu \right] 
\ \left({\rm for} \ \tau < t_{\rm s1}\right) \\
\\
\xi \exp \left[\eta / \tau^2 \right]
\ \left({\rm for} \ t_{\rm s1} \leq \tau < t_{\rm s2}\right)\\
\\
\left( \kappa / \sigma \right) \exp \left[ - \frac{\tau^2}{2\sigma^2} \right] 
\ \left({\rm for} \ t_{\rm s2} \leq \tau < t_{\rm s3}\right)\\
\\
\left( \kappa / \sigma \right) \exp \left[ \frac{t_{\rm s3} (t_{\rm s3} -2\tau)}{2\sigma^2} \right] 
\ \left({\rm for} \ \tau \geq t_{\rm s3}\right),
\end{array}
\right.
\end{align}
where $t_{{\rm s}i}$ indicate the range of each sub-function. The variable $\tau$ is defined as: $\tau \equiv t - t_{\rm c}$, where $t_{\rm c}$ is the peak center. Thus, $f_{\rm p}$ has seven independent parameters, not including the characteristic times $t_{{\rm s}i}$. The number of parameters can be analytically reduced to three by imposing the condition that continuity in the functions and their derivatives at each time $t_{{\rm s}i}$; the remaining independent parameters are then $t_{\rm c}$,$\kappa$, and $\sigma$. 

In addition to the primary peaks, we observed bump structures neighboring the intense peaks in sufficiently high-statistics TOF spectra. These bumps maintain a constant intensity and position relative to the primary peak and are deduced to be the result from the following process: secondary electrons are emitted from the surface of the MCP ion detector, they accelerate to nearby surfaces and produce tertiary ions which accelerate back to the MCP. Hence, these bumps are not actual events and depend only on the intensities of parent peaks. The bump shapes are modeled with the following Gaussian function,
\begin{equation}
f_{\rm b}(\tau) = \left( \kappa_{\rm b} / \sigma_{\rm b} \right) \exp \left[ - \left(\tau -t_{\rm b} \right)^2 / \left(2\sigma_{\rm b}^2 \right) \right].
\end{equation} 
In the fitting algorithm, $t_{\rm s1}$ and $t_{\rm s2}$ were determined by scaling the $\sigma$ parameter of mass reference peaks relative to that of $^{85}$Rb$^+$: $t_{\rm s1} = t_{\rm s1,85Rb} \times (\sigma/\sigma_{\rm 85Rb})$ and $t_{\rm s2} = t_{\rm s2,85Rb} \times (\sigma/\sigma_{\rm 85Rb})$. The $t_{\rm s3}$ values were determined as independent parameters in fitting the mass reference peaks, then fixed for each isobaric species of interest. The bump height parameter $\kappa_{\rm b}$ was calculated assuming a constant relative intensity, $\kappa_{\rm b} = \kappa_{\rm b, 85Rb} \times (\kappa/\kappa_{\rm 85Rb})$.  The other two bump parameters, $\sigma_{\rm b}$ and $t_{\rm b}$, were fixed based on $^{85}$Rb$^+$ fitted results.
 
For the species of interest, the only free parameters in the fitting function are the peak center $t_{\rm c}$ and the peak height $\kappa$. The $\sigma$ and $t_{\rm s3}$ parameters were determined from the mass reference values.  In the fitting algorithm, to enhance the mass precision, the $\tau$ parameter was treated as a function of $t$, $t_{\rm ref}$, and $\rho$, where $\rho$ is the TOF ratio from Eq.~\ref{eqSingleRef}. Then the fitting function $F$ for $N$ peaks is described by 
 \begin{align}
&&F(t, t_{\rm ref}, \rho_1, \cdots, \rho_{\rm N},\kappa_1, \cdots, \kappa_{\rm N}) = \hspace{4em} \nonumber  \\
&&\sum_{i=1}^{N} \{ f_{\rm p} (t, t_{\rm ref}, \rho_i,\kappa_i) + f_{\rm b} (t, t_{\rm ref}, \rho_i,\kappa_i)\}.
\end{align}
If the $j^{th}$ peak corresponds to the mass reference, $\rho_j$ is always unity; the total number of fitting parameters is $2 \times N$.

\section{Experimental accuracy}

In the intermediate-mass neutron-deficient region, the production mechanism typically provides a large range of isobaric species which can be simultaneously delivered to the MRTOF-MS.  We take advantage of this and utilize isobaric references within each isobaric chain to suppress mass-dependent systematic errors.  To minimize the possibility of introducing errors from inaccurately determined reference masses \citep{Schury2007}, we select the reference species as the highest intensity spectral peak corresponding to a nuclide whose mass value has been flagged as being derived from an ``absolute mass-doublet" measurement as defined in AME16 as well as having no known long-lived isomeric states.

Before we can hope to bring the MRTOF-MS to bear on such problems as CKM unitarity, we must have reasonable confidence in the accuracy limit of the device.  When using a simultaneously acquired isobaric mass reference, the accuracy is essentially limited by peak fitting. Primarily, there is a fundamental concern of how well the model function represents the real data.  In addition to that, we must be sure that we do not introduce any significant biases, \emph{e.g.}, in the process of binning the data.

To test for bias introduced in the process of binning the data, we use a typical $^{65}$Ga spectrum.  Following the same procedures as outlined above, we first determine the mass of $^{65}$Ga using a variety of bin sizes for the TOF data.  The variations introduced by differing bin sizes in the range of 0.8~ns to 12.8~ns are smaller than the 1.3$\times$10$^{-8}$ AME16 relative mass uncertainty. A second possibility for the introduction of a bias is the choice of time to start the binning. It is possible that some bias could be introduced based on the choice to begin the binning process {\emph e.g.}, at $t=6346500$~ns or $t=6346501$~ns. To investigate this, we used the same $^{65}$Ga spectrum, with 6.4~ns binning. The start of the binned spectrum was systematically shifted in increments of 0.8~ns, and the same procedures as previously described were used to determine the mass of $^{65}$Ga in each case. The fluctuation of the mass values was smaller than the uncertainty of the AME16 value and no effect on the mass value determined for $^{65}$Ga could be discerned. From these two studies, we confirmed that systematic effects in $^{65}$Ga measurements from binning are below the level of our precision.

It is worth noting that we have previously published results showing an accurate relative mass precision of 6$\times$10$^{-8}$ \citep{Schury2014} in the case of the well-resolved $^{40}$Ca$^+$/$^{40}$K$^+$ isobaric doublet. However in several of the measurements analyzed in this study, the peaks were not completely resolved.  In principle, it would be possible to use convoluted peak fitting in cases where peaks significantly overlap.  However, it has been the authors' experience that such fittings can be fraught with inaccurate results, particularly when relative intensities are not known $ab~initio$.

In some cases, there were known (or suspected) long-lived isomeric states at the $\sim$100~keV-level.  With our present mass resolving power, such isomeric states would be separated from the ground state for $\approx$20\% of FWHM.  In such a case, even an analytically-derived model function would achieve low-precision without knowledge of the isomeric population ratio.  As we lack knowledge of the isomeric population and utilize a phenomenologically-derived model function, we must designate the results from such cases as unreliable.

Even in better resolved cases, we must set some limit for requisite peak separation.  We consider two identical Gaussian peaks with some separation, with a fitting range chosen from -$\infty$ to the center of separation of the peaks.  If the peaks have FWHM separation, one can easily find that the fit result will be biased toward the rightmost peak by $\approx$11\% of FWHM.  If the peaks have full-width fifth-maximum separation, the bias falls to $\approx$3\%  of FWHM.  With full-width tenth-maximum (FWTM) separation, the bias falls to $\lesssim$0.5\%  of FWHM.  Thus, to ensure that peak overlap cannot bias the result beyond our desired level of precision, we also denote data wherein adjacent peaks were not resolved beyond the FWTM level to designate the mass derived from these data as possibly unreliable.

\section{Results}

Mass measurements were performed with two different settings of GARIS-II:  $B\rho = 0.86$~Tm and $B\rho = 1.01$~Tm, corresponding to reaction products from the sulfur targets ($A = 65-67$) and reactions products from the titanium backings ($A = 79-81$), respectively. Atomic and molecular ions of 25 species were identified within the TOF spectra of six different isobaric chains. 

\begin{figure*}[!b]
  \centering
  \includegraphics[width=1\textwidth, bb=0 0 720 252]{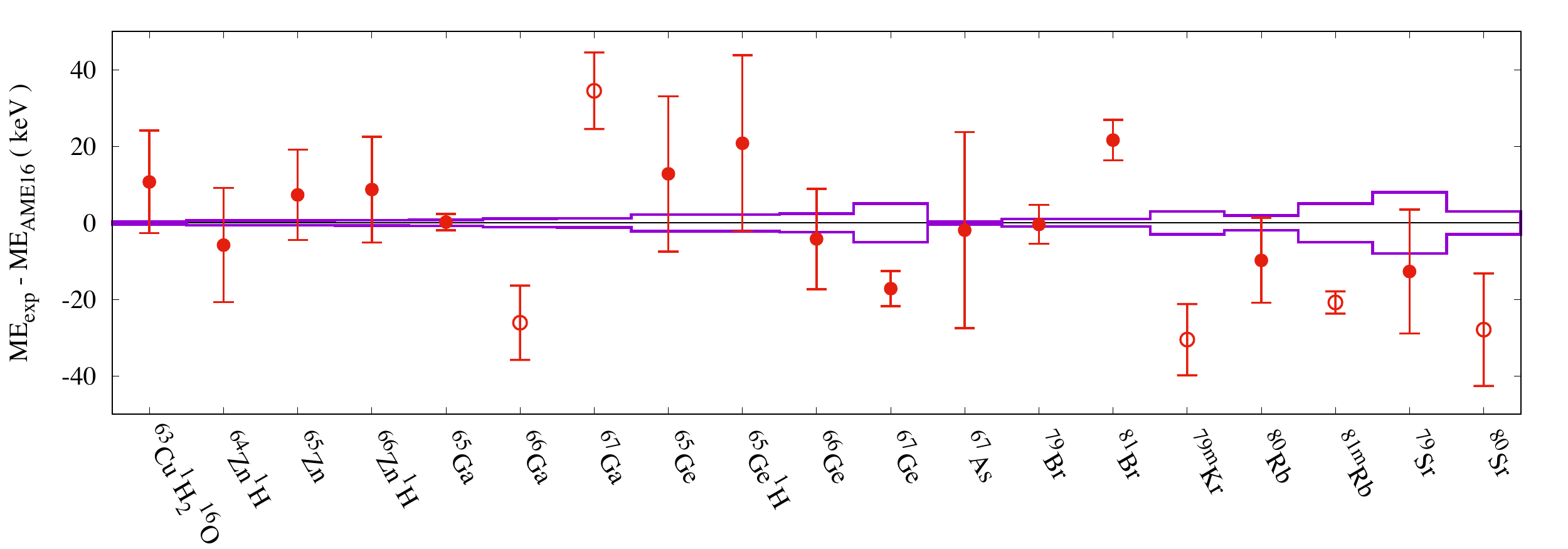}
  \caption{Differences between the present measurement results and the AME16 values. Purple lines represent errors of the AME16 values. The open symbols indicate data derived from spectral peaks insufficiently separated from adjacent spectral peaks. \label{summary_fig}}
   \vspace*{-10pt}
\end{figure*}

A summary of the results is shown in Table~\ref{summary} and Fig~\ref{summary_fig}. The relative precisions of  $\delta m/m = 4.1\times10^{-7}~{\rm to}~ 3.5\times10^{-8}$ were achieved in the present study. Details of the results for each $A/q$ series will be discussed in the following parts.

\begin{table*}
\begin{center}
\begin{threeparttable}
\caption{\label{tab:table1}The squares of time-of-flight ratio $\rho^2$, which are equivalent to the mass ratio, and the mass excess values ME$_{\rm exp}$ in the present study. The nuclides used as atomic mass references are shown in column ``Ref.". Label "$_{\rm L}$" indicates values deduced by employing likelihood method. ME$_{\rm lit}$ indicates mass excess values from AME16 and $\Delta$ME represents the differences between mass excess values found in AME16 and those of the present study: $\Delta{\rm ME} \equiv {\rm ME}_{\rm exp} - {\rm ME}_{\rm lit}$. The $\delta$m/m column provides relative mass precisions of the present measurements. The brackets in the $\Delta$ME column denote measurements that do not satisfy the reliability conditions as described in the text. The bracket shape designates which criterion was not satisfied: $\langle \cdots \rangle$ denotes contamination with unresolvable isomers and $[ \cdots ]$ indicates the undue influence of intense neighboring peaks. \label{summary}}
\renewcommand{\arraystretch}{1.2}
\renewcommand\thefootnote{\alph{footnote}}
\begin{tabular}{cclllcc} \hline
\textrm{Species}&
\textrm{Ref.}&
\multicolumn{1}{c}{\textrm{$\rho^2$}}&
\multicolumn{1}{c}{\textrm{ME$_{\rm exp}$ (keV)}}&
\multicolumn{1}{c}{\textrm{ME$_{\rm lit}$ (keV)}}&
\multicolumn{1}{c}{\textrm{$\Delta$ME (keV)}}&
\multicolumn{1}{c}{\textrm{$\delta$m/m}}\\ \hline
$^{63}$Cu$^{1}$H$_2$$^{16}$O$^{+}$&$^{81}$Sr$^{+}$&1.00020961(17)$_{\rm L}$&\hspace{0.7em}$-$55728(13)&
$-$55738.9(4)&11(13)&$1.8\times10^{-7}$\\
\hline
$^{64}$Zn$^{1}$H$^{+}$&$^{65}$Cu$^{+}$&1.00014125(25)&\hspace{0.7em}$-$58721(15)&$-$58715.0(6)&$-$6(15)&$2.5\times10^{-7}$\\
$^{65}$Zn$^{+}$&$^{65}$Cu$^{+}$&1.00002247(20)&\hspace{0.7em}$-$65905(12)&$-$65912.0(6)&
7(12)&$2.0\times10^{-7}$\\
$^{66}$Zn$^{1}$H$^{+}$&$^{67}$Zn$^{+}$&1.00010072(22)&\hspace{0.7em}$-$61602(14)&$-$61610.2(7)&9(14)&$2.2\times10^{-7}$\\
 \hline
$^{65}$Ga$^{+}$&$^{65}$Cu$^{+}$&1.000076165(33)&\hspace{0.7em}$-$62657.1(21)&$-$62657.5(8)&0.2(22)&$3.5\times10^{-8}$\\
$^{66}$Ga$^{+}$&$^{66}$Zn$^{+}$&1.00008385(16)$_{\rm L}$&\hspace{0.1em}[ $-$63749.8(97) ]&$-$63723.7(11)&[ $-$21.6(97) ]&$1.6\times10^{-7}$\\
$^{67}$Ga$^{+}$&$^{67}$Zn$^{+}$&1.00001662(16)&\hspace{0.1em}[ $-$66845(10) ]&$-$66879.0(12)&[ 35(10) ]&$1.6\times10^{-7}$\\
\hline
$^{65}$Ge$^{+}$&$^{65}$Cu$^{+}$&1.00017854(34)&\hspace{0.7em}$-$56465(20)&$-$56478.2(22)&13(20)&$3.4\times10^{-7}$\\
$^{65}$Ge$^{1}$H$^{+}$&$^{66}$Zn$^{+}$&1.00032130(37)&\hspace{0.7em}$-$49168(23)&$-$49189.2(22)&21(23)&$3.8\times10^{-7}$\\
$^{66}$Ge$^{+}$&$^{66}$Zn$^{+}$&1.00011868(21)$_{\rm L}$&\hspace{0.7em}$-$61611(13)&$-$61607.0(24)&$-$4(13)&$2.1\times10^{-7}$\\
$^{67}$Ge$^{+}$&$^{67}$Zn$^{+}$&1.000083493(73)&\hspace{0.7em}$-$62675.2(46)&$-$62658(5)&$-$17(7)&$7.4\times10^{-8}$\\
\hline
$^{67}$As$^{+}$&$^{67}$Zn$^{+}$&1.00018112(41)$_{\rm L}$&\hspace{0.7em}$-$56589(26)&$-$56587.2(4)&
$-$2(26)&$4.1\times10^{-7}$\\
\hline
$^{79}$Br$^{+}$&$^{79}$Rb$^{+}$&0.999928378(63)&\hspace{0.7em}$-$76068.4(51)&$-$76068.0(10)&$-$0.4(52)&$7.0\times10^{-8}$\\
$^{81}$Br$^{+}$&$^{81}$Sr$^{+}$&0.999914739(60)&\hspace{0.7em}$-$77955.4(53)&$-$77977.0(10)&21.6(54)&$7.0\times10^{-8}$\\
\hline
$^{79}$Kr$^{+}$
&$^{79}$Rb$^{+}$&0.99995185(12)\tnote{a}&\hspace{0em}$\langle$ $-$74342.7(93) $\rangle$&$-$74442(3)&$\langle$ 100(10) $\rangle$&$1.3\times10^{-7}$\\
\hline
$^{80}$Rb$^{+}$&$^{80}$Kr$^{+}$&1.00007668(15)&\hspace{0.7em}$-$72185(11)&$-$72175.5(19)&$-$10(11)&$1.5\times10^{-7}$\\
$^{81}$Rb$^{+}$
&$^{81}$Sr$^{+}$&0.999948753(11)\tnote{b}&\hspace{0em}$\langle$ $-$75391.5(29) $\rangle$&$-$75457(5)&$\langle$ 65(6) $\rangle$&$3.9\times10^{-8}$\\
\hline
$^{79}$Sr$^{+}$&$^{79}$Rb$^{+}$&1.00007227(22)$_{\rm L}$&\hspace{0.7em}$-$65490(16)&$-$65477(8)&
$-$13(18)&$2.2\times10^{-7}$\\
$^{80}$Sr$^{+}$&$^{80}$Kr$^{+}$&1.00010148(20)&\hspace{0.1em}[ $-$70339(15) ]&$-$70311(3)&[ $-$28(15) ]&$2.0\times10^{-7}$\\ \hline
\end{tabular}

\begin{tablenotes}
\item[a]{given for admixture with the isomer of $E_{\rm X} = 129.77$ keV, $T_{1/2} = 50$ sec \citep{Singh2002}}
\item[b]{given for admixture with the isomer of $E_{\rm X} = 86.31$ keV, $T_{1/2} = 30.5$ min \citep{Baglin2008}}
\end{tablenotes}

\end{threeparttable}
\end{center}
\end{table*}

\begin{figure*}
  \centering
  \includegraphics[width=1\textwidth, bb=0 0 842 595, clip, trim=40 100 30 110]{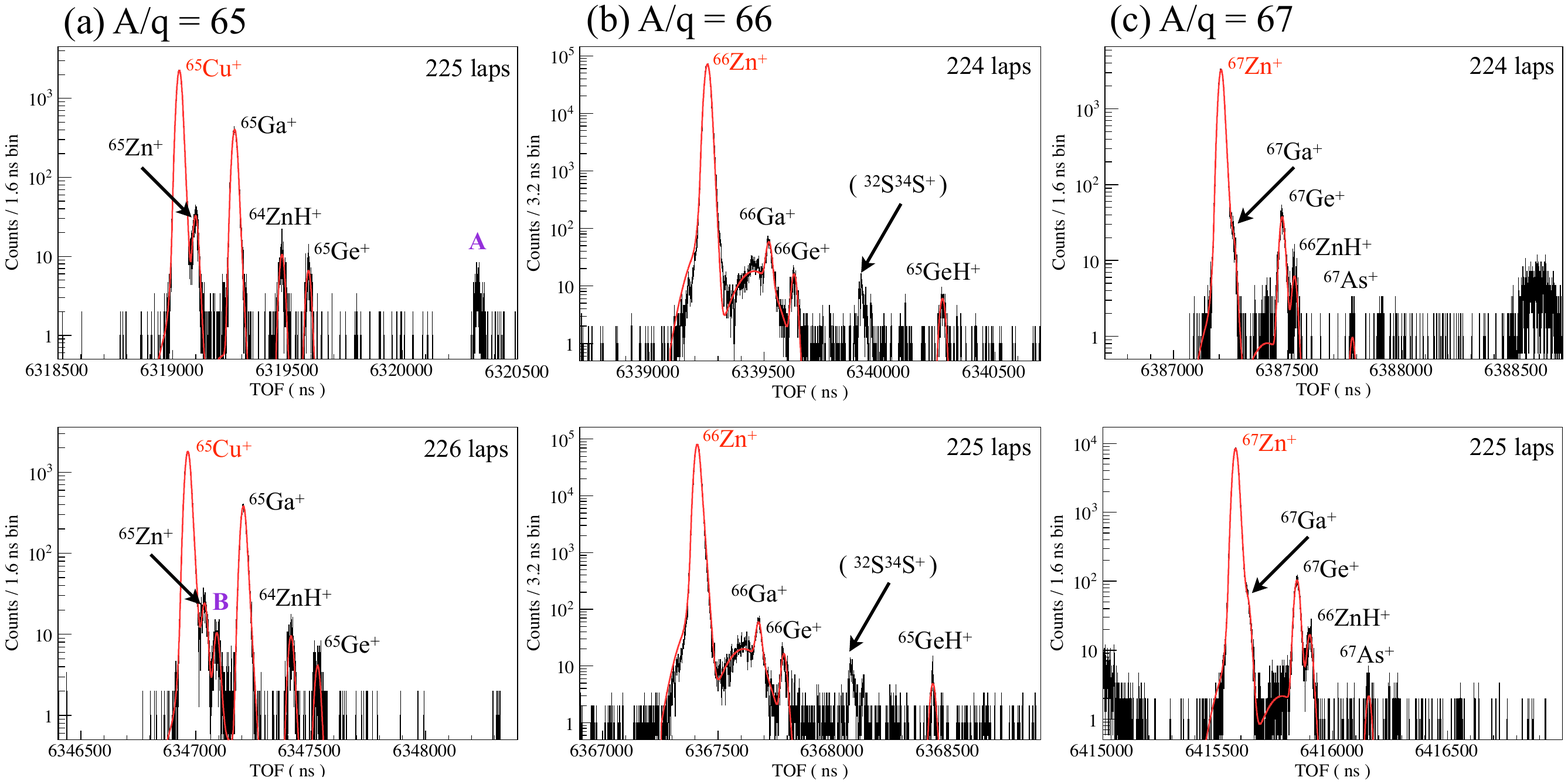}
  \caption{(a) Time-of-flight spectra for $A/q = 65$ at 225 and 226~laps. Red lines indicate the fit. The nuclide which is used as the mass reference is indicated by red characters. The peaks labeled ``A" and ``B" are transient contaminants, making a number of laps different from $A/q = 65$ ions.  This can be inferred the apparent movement of the peak between $N_{65}$=225~laps and $N_{65}$=226~laps. Unlabelled peaks in subsequent figures can be presumed to be of similar origin. (b) Time-of-flight spectra of $A/q = 66$ at 224 and 225 laps. (c) Time-of-flight spectra of $A/q = 67$ at 224 and 225 laps. \label{A65-67}}
  \vspace*{-10pt}
\end{figure*} 

\begin{figure*}
  \centering
  \includegraphics[width=1\textwidth, bb=0 0 842 595, clip, trim=40 100 30 110]{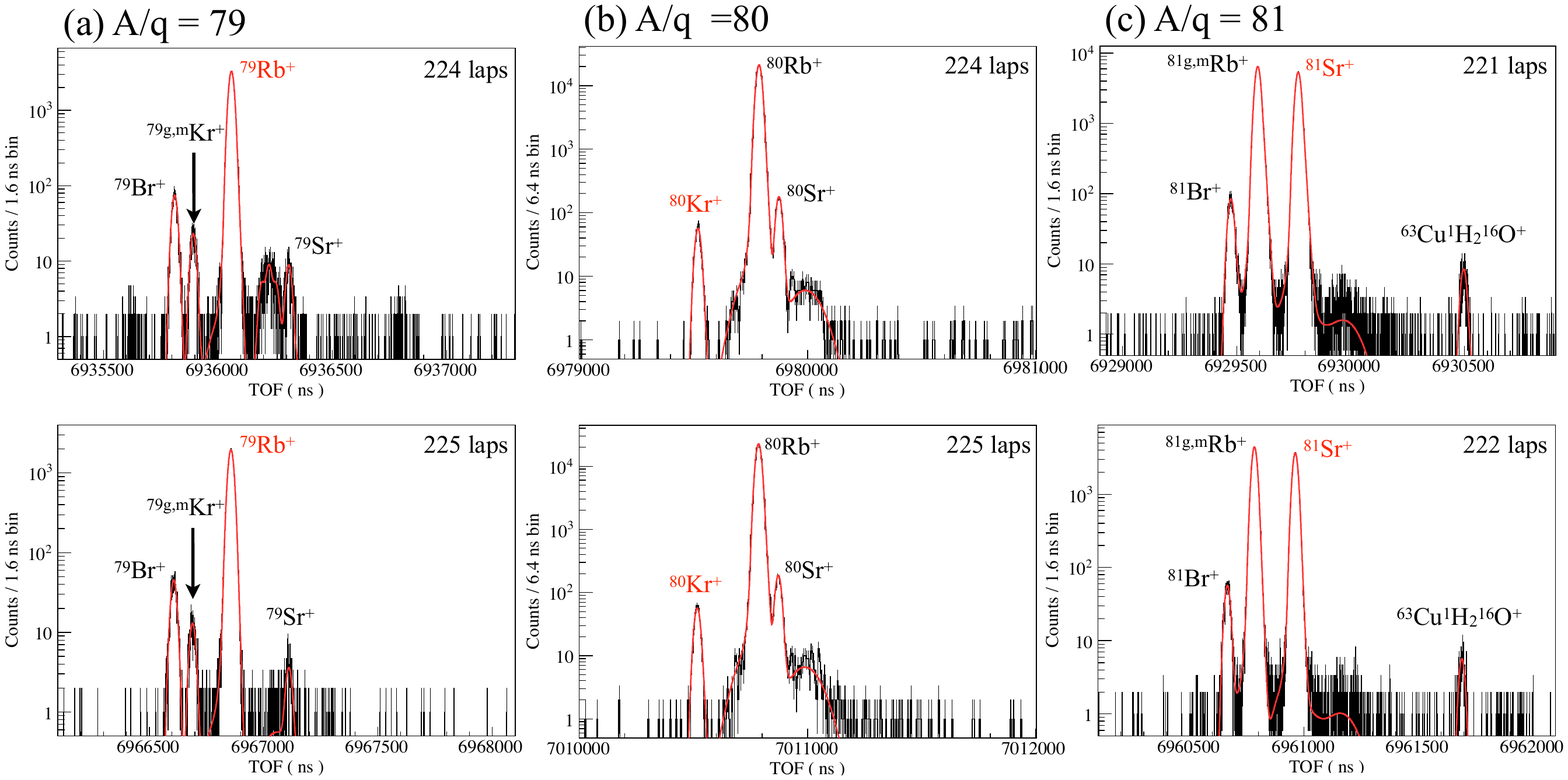}
  \caption{(a) Time-of-flight spectra for $A/q = 79$ at 224 and 225~laps. (b) Time-of-flight spectra of $A/q = 80$ at 224 and 225 laps. (c) Time-of-flight spectra of $A/q = 81$ at 221 and 222 laps. See Fig.~\ref{A65-67} for details. \label{A79-81}}
  \vspace*{-10pt}
\end{figure*} 

\subsection{${\bf A/q = 65}$}
For the $A/q=65$ series, mass measurements were performed with three different lap values (225-227~laps). Typical TOF spectra are shown in Fig.~\ref{A65-67} -- (a). $^{65}$Cu was employed as the mass reference. All mass excess values in this series were in agreement with the AME16 values. For $^{65}$Ga, a mass uncertainty of 2.1 keV, corresponding to a relative precision of $\delta m / m = 3.5\times10^{-8}$, was obtained. This is the most precise measurement yet performed by MRTOF-MS. The accumulated number of $^{65}$Ga events was $1.9 \times 10^{4}$. 

\subsection{${\bf A/q = 66}$}
Mass excess values of the $A/q=66$ series were obtained from measurements with six different lap values (223-227, 299~laps). Here, $^{66}$Zn was used as the mass reference. A discrepancy with the AME16 value was found for $^{66}$Ga. However, from Fig.~\ref{A65-67} --(b) it can be seen that $^{66}$Ga peaks are located in the bump structures of $^{66}$Zn peaks and they do not satisfy the FWTM reliability requirement. The mass of the peak located between $^{66}$Ge$^{+}$ and $^{65}$GeH$^{+}$ is consistent with the sulfur-dimer $^{32}$S$^{34}$S$^{+}$, however, the source of such a molecule is unclear and only a tentative assignment is made. 

\subsection{${\bf A/q = 67}$}
In the $A/q=67$ series, four measurements (223-225, 227~laps) were performed. The mass reference in this series was $^{67}$Zn. The mass values of two nuclides, $^{67}$Ga and $^{67}$Ge, were inconsistent with the AME16 values. It is seen in Fig.~\ref{A65-67} -- (c) that the $^{67}$Ga peaks are fully embedded in the $^{67}$Zn peaks, eliminating $^{67}$Ga from further discussion. In contrast to the case of $^{67}$Ga, the $^{67}$Ge peaks are well-resolved. The discrepancy in the $^{67}$Ge mass value was seen to be 17~keV, corresponding to 2.4-$\sigma$. The $^{67}$Ge mass value of AME16 was evaluated by an indirect method, namely threshold measurements of the $^{64}{\rm Zn}( \alpha,{\rm n})^{67}{\rm Ge}$ reaction \citep{Murphy1978,Al-Naser1979}. The threshold values of $^{64}{\rm Zn}( \alpha,{\rm n})^{67}{\rm Ge}$ were extrapolated by using the 2/3-law \cite{Bondelid1964}, wherein the deexcitation $\gamma$-ray yield $y$ is correlated to the induced $\alpha$-particle energy $E_{\alpha}$ by $y^{2/3} \propto E_{\alpha}$.  However, when the Coulomb barrier height exceeds the reaction Q-value, the yield curve can be expressed as
\begin{equation}
y(E_{\alpha}) \propto E_{\alpha} \int_{0}^{E_{\alpha}} d\varepsilon \ \frac{1}{\sqrt{\varepsilon}} \exp(-\frac{V_{\alpha, 64Zn}}{\sqrt{\varepsilon+E_{th}}}),
\end{equation}
which is distorted from the 2/3-law. This distorted yield curve leads the overestimation of the reaction threshold. The discrepancy of $^{67}$Ge mass value is resolved by such consideration of the influence of Coulomb barrier. Thus a new mass excess value, ${\rm ME} = -62675.2(46)$~keV, for $^{67}$Ge is proposed

\subsection{${\bf A/q = 79}$}
For the A/q=79 series as show in Fig~\ref{A79-81} -- (a), $^{79}$Rb was taken as the mass reference since it is the highest intensity nuclide in the set which has previously been evaluated in a Penning trap measurement \citep{Kellerbauer2007} and has no known long-lived isomeric states. There are five measurements in this series (224,225, 227-229~laps). The mass values obtained for $^{79}$Br and $^{79}$Sr are in good agreement with those in AME16. The discrepancy of the $^{79}$Kr mass value is set aside due to an unresolved, long-lived isomeric state. 

\subsection{${\bf A/q = 80}$}
There are two measurements with different lap values (224 and 225 laps) for the $A/q=80$ series as show in Fig~\ref{A79-81} -- (b).  $^{80}$Kr was taken as the mass reference since it is the only nuclide which satisfies the requirements to be a mass reference. The mass value of $^{80}$Sr in the present study is inconsistent with the AME16 value. However, as can be seen in Fig.~\ref{A79-81} -- (b), the FWTM reliability requirement is not met. 

\subsection{${\bf A/q = 81}$}
In the A/q = 81 series, mass excess values were determined by four measurements with different lap values (221, 222, 224, and 225~laps). Fig~\ref{A79-81} -- (c) shows the measured TOF spectra of the A/q = 81 series. $^{81}$Sr was selected as a mass reference for the $A/q=81$ series since its mass value was determined by Penning trap measurements \citep{Otto1994,Haettner2011} and it also provides high-intensity peaks. In addition, $^{81}$Sr has no known long-lived isomeric states. In this series inconsistent mass values were found for two nuclides: $^{81}$Rb and $^{81}$Br. For $^{81}$Rb we dismiss the discrepancy as being due to the admixture of an unresolved, long-lived isomeric state. The $^{81}$Br peak, however, satisfies the reliability requirements.  As in the case of $^{67}$Ge, the mass value of $^{81}$Br was evaluated using the results of indirect measurements connecting to the absolute mass-doublet nuclide $^{82}$Kr: $^{81}{\rm Br}({\rm n},\gamma) ^{82}{\rm Br} (\beta^{-}) ^{82}{\rm Kr}$. We claim that the connection between $^{82}{\rm Br}$ and $^{82}{\rm Kr}$ is susceptible to error since it depends on a $\beta$-decay endpoint measurement \citep{Waddell1956}.   The excitation energies of $^{82}{\rm Kr}$ adopted by \citep{Waddell1956}, which were measured by a NaI scintillator, have slightly smaller values ($\sim$10 keV) as compared with currently adopted ones measured with Ge detectors \citep{Tuli2003}.  If the modern excitation energies are used, the resultant $Q_{\beta}$-value agrees with the presently measurement mass of $^{81}$Br. Thus a new mass excess value for $^{81}{\rm Br}$ of ${\rm ME} = -77955.4(53)$~keV is proposed.

\section{Discussion}

\begin{figure}[!t]
  \centering
  \includegraphics[width=0.5\textwidth, clip, bb=0 0 567 386]{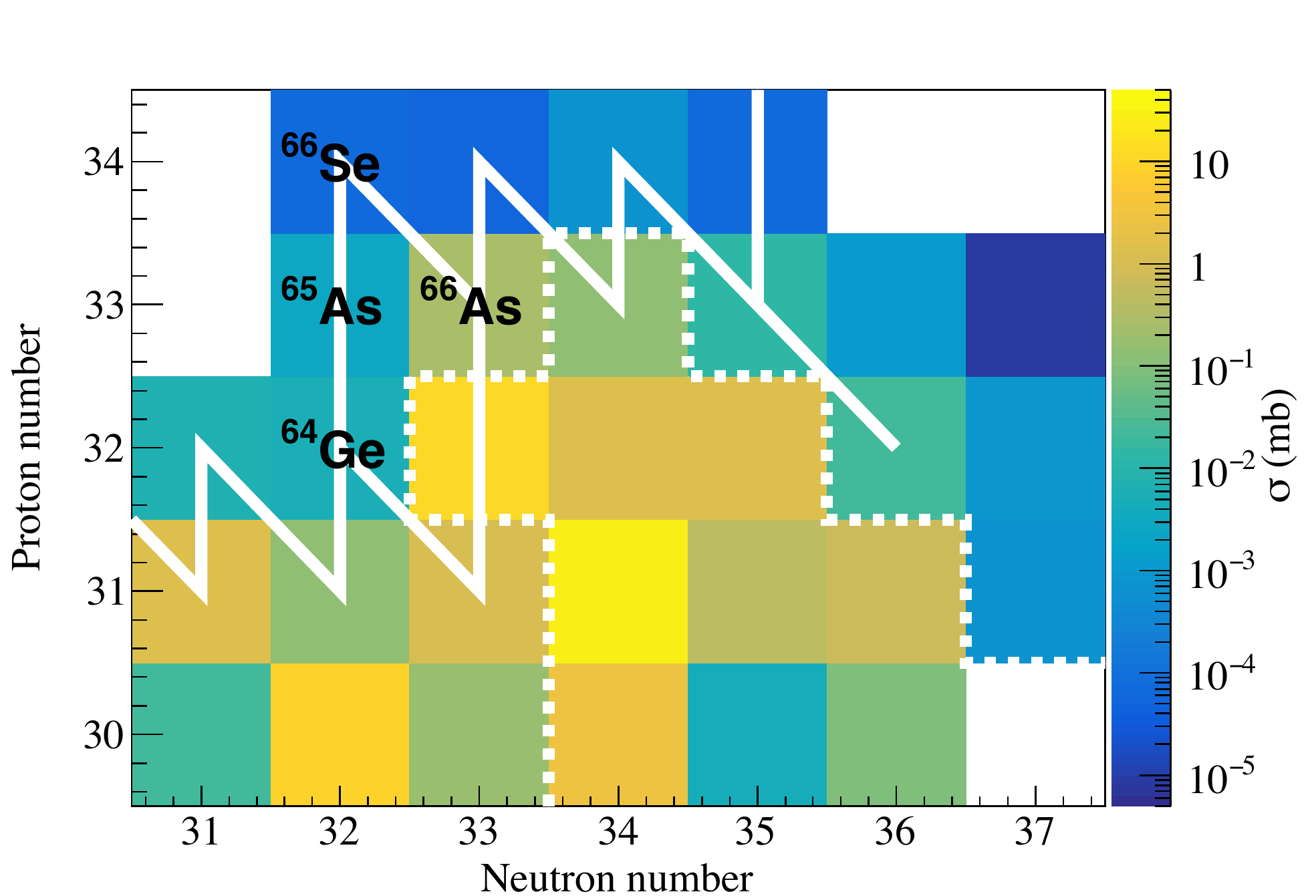}
  \caption{Predicted production cross-sections of the fusion-evaporation reaction $^{\rm nat}{\rm S}$($^{36}{\rm Ar}$,X). The white dotted line shows the boundary of nuclides whose masses were measured in the present study. The white bold line represents the $rp$-process pathway branches that have more than 10\% fraction of mass flow \citep{Schatz2001}. \label{CS}}
\end{figure} 

Figure \ref{CS} shows the theoretical production cross-sections of $^{\rm nat}{\rm S}$($^{36}{\rm Ar}$,X) reactions in the present measurement setting. The LisFus code \citep{Tarasov2003} was used to calculate these values. In the present study it was possible to access nuclides with cross-sections slightly less than one millibarn. The desired nuclides, however, are predicted to have cross-sections that are one to three orders of magnitude smaller than those measured in the present study (see Fig. \ref{CS}). 

The total efficiency of the SHE-mass facility in the present measurements could be estimated by counting the number of $^{80}$Rb events in both the $\beta$-activity counter and the MRTOF-MS. An intensity of $^{80}$Rb ions at the GARIS-II focal plane was estimated to be $1 \times 10^{6} \ {\rm cps \cdot p\mu A^{-1} }$ from $\beta$-activity counting rate and the detection efficiency calculated by GEANT4 simulations \citep{Agostinelli2003}. The counting rate at the MRTOF-MS was  $4 \times 10^{2} \ {\rm cps \cdot p\mu A^{-1} }$. The total efficiency of the SHE-mass facility excluding the transmission efficiency of GARIS-II  for the present measurements was estimated to be $\sim$0.03\%.

In the present series of measurements, only singl-charged ions have been studied in the MRTOF-MS spectra. A recent gas-cell study suggests that most elements could be extracted as doubly-charged ions \citep{Schury2017b}. This implies that the single-charged state could be a very minor component among all ionization states in the gas-cell and provides room for great improvement in the total efficiency.  In the case of super-heavy element mass measurements at the SHE-mass facility, ions were extracted as doubly-charged ions and the total efficiencies reached the few percent level, which is roughly one hundred times higher than the present value. The mass measurements of the Md isotopes, which have production cross-sections on the order of 100~nb, have been achieved by optimizing the SHE-mass facility for doubly-charged ions \citep{Ito2017submitted}. This improved total efficiency could allow to access nuclides with microbarn or sub-microbarn, production cross sections. The SHE-mass facility can be used for not only the study of super-heavy elements but also the intermediate-mass nuclides in more neutron-deficient side, which are crucial to the $rp$-process.

\section{Summary and Conclusions}

In conclusion, the masses of $^{63}$Cu, $^{64-66}$Zn, $^{65}$Ga, $^{65-67}$Ge, $^{67}$As, $^{78,81}$Br, $^{80}$Rb, and $^{79}$Sr  were measured using the MRTOF-MS combined with GARIS-II under the minimal B$\rho$-value difference of the primary beam and reaction products. The masses of these nuclides have been determined by the single reference method using known isobaric references. There are some inconsistencies between AME16 values, and two new mass excess values are proposed: ${\rm ME (^{67}Ge)} = -62675.2(46)$~keV and ${\rm ME (^{81}Br)} = -77955.4(53)$~keV. This result reinforces the need for direct mass measurements of all nuclides, even for stable isotopes, if their masses were previously evaluated by indirect techniques. The relative mass precisions in the present study span the range from  $\delta m/m = 4.1\times10^{-7}~{\rm to}~ 3.5\times10^{-8}$. In the most precise measurement, that of $^{65}$Ga, a mass uncertainty of 2.1~keV was obtained. This result shows that mass measurements satisfying the requirement of the CKM matrix can be achieved with the MRTOF-MS, given sufficient statistics. The SHE-mass facility is seen to be suitable for precision mass measurements of intermediate-mass, more neutron-deficient nuclides, crucial to the $rp$-process. \\

We would like to express our sincere gratitude to the RIKEN Nishina Center for Accelerator-Based Science and the Center for Nuclear Science at the University of Tokyo for their support of the present measurements. H. S. acknowledges support from the US National Science Foundation under PHY-1565546 and PHY-1430152 (JINA Center for the Evolution of the Elements). This study was supported by the Japan Society for the Promotion of Science KAKENHI, Grant Number 24224008, 15H02096, 15K05116, and 17H06090.

\bibliography{p-rich_mass}

\end{document}